\documentclass[conference]{IEEEtran}
\IEEEoverridecommandlockouts

\usepackage[dvipsnames]{xcolor}
\usepackage[utf8]{inputenc} 
\usepackage{comment}
\usepackage[T1]{fontenc}
\usepackage{url}
\usepackage{ifthen}
\usepackage{cite}
\usepackage{amsmath}
\usepackage{mathtools,cuted}
\usepackage{amssymb} 
\usepackage{cite}
\usepackage{pgf,tikz,pgfplots}
\usepackage{algorithmic}
\usepackage{graphicx}
\usepackage{tabularx}
\usepackage{multirow}
\usepackage{array}
\usepackage{tikz}
\usepackage{ctable}
\usepackage{stfloats}

\newcommand{\otoprule}{\midrule[\heavyrulewidth]}

\makeatletter
\renewcommand\env@matrix[1][\arraystretch]{%
  \edef\arraystretch{#1}%
  \hskip -\arraycolsep
  \let\@ifnextchar\new@ifnextchar
  \array{\c@MaxMatrixCols c}}
\makeatother
\newcommand{\w}{w}  
\newcommand{\rate}{\Omega}

\newcommand{\ceil}[1]{\left\lceil #1 \right\rceil}

\newcommand{\bx}{\boldsymbol{x}}
\newcommand{\bA}{\boldsymbol{A}}
\newcommand{\bB}{\boldsymbol{B}}
\newcommand{\bs}{\boldsymbol{s}}
\newcommand{\bU}{\boldsymbol{U}}
\newcommand{\bH}{\boldsymbol{H}_\text{t}}
\newcommand{\ns}{n_\mathsf{u}}
\newcommand{\dv}{d_\mathsf{v}}
\newcommand{\dc}{d_\mathsf{c}}
\newcommand{\oneb}{\boldsymbol{1}_{1\times \dc}}

\newcommand{\ga}{\gamma}

\newcolumntype{M}[1]{>{\centering\arraybackslash}p{#1}}
\newcommand{\pwl}{^{(\ell)}}
\newcommand{\pwlp}{^{(\ell-1)}}

\newcommand{\qz}{q^{(\ell)}_0}
\newcommand{\qd}{q^{(\ell)}_1}
\newcommand{\qzm}{q^{(\ell-1)}_0}
\newcommand{\qdm}{q^{(\ell-1)}_1}
\newcommand{\pz}{p^{(\ell)}_0}
\newcommand{\pd}{p^{(\ell)}_1}
\newcommand{\pzm}{p^{(\ell-1)}_0}
\newcommand{\pdm}{p^{(\ell-1)}_1}
\newcommand{\qzt}{q^{(\ell)}_{0,\tau}}
\newcommand{\qdt}{q^{(\ell)}_{1,\tau}}
\newcommand{\qztm}{q^{(\ell-1)}_{0,\tau+j}}
\newcommand{\qdtm}{q^{(\ell-1)}_{1,\tau+j}}
\newcommand{\pzt}{p^{(\ell)}_{0,\tau}}
\newcommand{\pdt}{p^{(\ell)}_{1,\tau}}
\newcommand{\pztm}{p^{(\ell-1)}_{0,\tau-j}}
\newcommand{\pdtm}{p^{(\ell-1)}_{1,\tau-j}}

\makeatletter
\newcommand{\ostar}{\mathbin{\mathpalette\make@circled\star}}
\newcommand{\make@circled}[2]{%
  \ooalign{$\m@th#1\smallbigcirc{#1}$\cr\hidewidth$\m@th#1#2$\hidewidth\cr}%
}
\newcommand{\smallbigcirc}[1]{%
  \vcenter{\hbox{\scalebox{0.77778}{$\m@th#1\bigcirc$}}}%
}
\makeatother

\definecolor{darkblue}{rgb}{0.07843,0.16863,0.54902}
\definecolor{darkgreen}{rgb}{0,0.49804,0}%
\definecolor{brown}{rgb}{0.85098, 0.32941, 0.10196}%

\newcommand{\transpose}{^\mathsf{T}}
\newcommand{\msfv}{\mathsf{v}}
\newcommand{\msfc}{\mathsf{c}}
\newcommand{\msfB}{\mathsf{B}}
\newcommand{\dcell}{d_\mathsf{c}^{(\ell)}}
\newcommand{\bsell}{\boldsymbol{s}^{(\ell)}}
\newcommand{\Bino}{\mathsf{Bino}}
\newcommand{\gammath}{\gamma_\mathsf{th}}
\newcommand{\Omegath}{\Omega_\mathsf{th}}
\newtheorem{proposition}{Proposition}

\begin{document}
\title{Low-Density Parity-Check Codes and Spatial Coupling for Quantitative Group Testing} 

\author{
\IEEEauthorblockN{Mgeni Makambi Mashauri\IEEEauthorrefmark{1}, Alexandre Graell i Amat\IEEEauthorrefmark{2}, and Michael Lentmaier\IEEEauthorrefmark{1}}
\IEEEauthorblockA{\IEEEauthorrefmark{1}Department of Electrical and Information Technology, Lund University, Lund, Sweden}
\IEEEauthorblockA{\IEEEauthorrefmark{2}Department of Electrical Engineering, Chalmers University of Technology, Gothenburg, Sweden}\\\vspace*{-1.5cm}

\thanks{This work was supported in part by the Excellence Center at Linköping-Lund in Information Technology (ELLIIT). The simulations were partly performed on resources provided by the Swedish National Infrastructure for Computing (SNIC) at center for scientific and technical computing at Lund University (LUNARC).}
}

\maketitle

\begin{abstract}
A non-adaptive quantitative group testing (GT) scheme based on sparse codes-on-graphs in combination with low-complexity peeling decoding was introduced and analyzed by Karimi \emph{et al.}. In this work, we propose a variant of this scheme based on low-density parity-check codes where the BCH codes at the constraint nodes are replaced by simple single parity-check  codes. Furthermore, we apply spatial coupling to both GT schemes, perform a density evolution analysis, and compare their  performance with and without coupling. Our analysis shows that both schemes improve with increasing coupling memory, and for all considered cases, it is observed that the LDPC code-based scheme substantially outperforms the original scheme. Simulation results for finite block length confirm the asymptotic density evolution thresholds.
\end{abstract}

\vspace*{-1.2mm}
\section{Introduction}
The general goal of group testing (GT) \cite{Dor43} is to identify the set of $k$ defective items among a population of $n$ items by efficiently pooling groups of items in order to reduce the total number of required tests $m < n$. 
In the sub-linear regime \cite{Gebhard19}, where the prevalence $\gamma=k/n$ tends to zero as $n$ increases, it has been demonstrated that sparse codes-on-graphs \cite{Wad13}, can identify all defective items with high probability with low-complexity iterative (peeling) decoding \cite{Saffron, Vem17}. In \cite{KEK2019R} and \cite{KEK2019}, this idea was extended  from non-quantitative to quantitative GT using $t$-error correcting BCH codes at the constraint nodes of a generalized low-density parity-check (GLDPC) code with regular and irregular variable node (VN) degrees, respectively. It turns out that the strongest codes, with largest VN degree $\dv$ and decoding radius $t$, do not  perform best with iterative decoding. Instead, the minimum number of required tests in  \cite{KEK2019R,KEK2019} is  achieved for $t=2$ and the distribution of $\dv$ has to be chosen carefully for every $t$. 

Spatial coupling of regular graphs is an attractive alternative to the sensitive optimization of irregular graphs, thanks to the threshold saturation phenomenon that leads to robust performance with iterative decoding even for large $\dv$. First observed for low-density parity-check (LDPC) codes \cite{LSC2010, Kud2011}, this behavior extends to other graph-based systems such as GLDPC codes \cite{MDOL2021} or iterative decoding and detection \cite{Ngu2012, MIL2021}. 
To the best of our knowledge, however, the concept of spatial coupling has never been applied to GT schemes.

Our main contribution in this paper is two-fold: first, we propose a novel quantitative GT scheme based on LDPC codes as an alternative to the GLDPC code-based GT scheme in \cite{KEK2019R,KEK2019}. A corresponding peeling decoder is presented, which cannot rely on local error correction of the component codes (since $t=0$) but instead takes advantage of the cases where either all or none of the items within a test are defective.  Second, we apply spatial coupling to both schemes. We further  perform a density evolution analysis of the LDPC code-based GT scheme and of the coupled schemes to investigate the effect of increasing coupling memory  for various combinations of $d_v$ and $t$.
We consider two scenarios for evaluating the schemes. In the first scenario, we fix the proportion of defective items $\gamma$ (prevalence) and compute the minimum required rate $\Omega$, defined as the number of tests per item. This allows for a comparison with the results presented in \cite{KEK2019R}.
In the second scenario, in order to study threshold saturation, we consider a fixed graph structure with rate $\Omega$ and analyze how much $\gamma$ can be increased while still maintaining reliable recovery of the items. For both scenarios, it can be observed that spatial coupling improves the performance as the coupling memory $w$ increases. In particular, the best thresholds $\gammath$ are  achieved for larger values of $\dv$.
Remarkably, the density evolution analysis also shows that the proposed LDPC code-based GT scheme  significantly outperforms the GLDPC code-based GT scheme of \cite{KEK2019R,KEK2019}. This is also true for the coupled schemes. Finally, we present finite block length simulation results for the LDPC code-based and GLDPC code-based GT schemes  that confirm the behavior observed in the asymptotic analysis.

\section{System Model}\label{sec:model}

We consider a population of $n$ items, each of which is defective with probability $\gamma$, referred to as the \emph{prevalence}. We represent the $n$ items by a binary vector $\bx=(x_1,\ldots,x_n)$, where $x_i=1$ if item $i$ is defective and $x_i=0$ if it is not. Vector $\bx$ is unknown and the goal of the GT scheme is to infer it. 

The GT consists of $m$ tests and can be represented by an $m\times n$ test matrix $\bA=(a_{i,j})$, where row $i$ corresponds to test $i$, column $j$ corresponds to item $j$, i.e., $x_j$, and $a_{i,j}=1$ if item $j$ participates in test $i$ and $a_{i,j}=0$ otherwise.

Here, we consider noiseless, non-adaptive quantitative group testing, where the result of each test correctly gives the number of defective items in the test.  The result of the $i$-th test, denoted by $s_i$, is therefore given by
\begin{align*}
 s_i=\sum_{j=1}^{n} x_ja_{i,j}\,.
\end{align*} 
We collect the results of the $m$ tests in the \emph{syndrome} vector $\bs=(s_1,\ldots,s_m)$. It holds
\begin{align*}
\bs=\bx\bA\transpose\,.
\end{align*} 
Based on the syndrome, the goal of GT is to estimate $\bx$ via a decoding operation.

The assignment of items to tests can be conveniently represented by a bipartite graph consisting of $n$ variable nodes (VNs) corresponding to the $n$ items and $m$ constraint nodes (CNs) corresponding to the $m$ tests. An edge between VN $j$, $\msfv_j$, and CN $i$, $\msfc_i$, is drawn if item $x_j$ participates in test $i$, i.e., if $a_{i,j}=1$.

Fig.~\ref{fig:bipartitegraph} shows the bipartite graph corresponding to a scenario with $6$ items and $3$ tests with assignment matrix
\begin{align}
\label{eq:Matrix}
\bA=\left(\begin{array}{cccccc}
    1 & 1 & 0 & 1 & 0 & 1\\
    0 & 1 & 1 & 1 & 1 & 0\\
    1 & 0 & 1 & 0 & 1 & 1
\end{array}\right) \ .
\end{align}

 The bipartite graph representation of quantitative GT traces a connection with codes-on-graphs. Hence, the theory of codes-on-graphs can be used to design good test matrices $\bA$ and analyze their properties.

\vspace*{-2mm}
\section{Preliminaries: Group Testing\\Based on GLDPC Codes}

\begin{figure}[t!]
\centerline{\includegraphics[width=0.75\columnwidth]{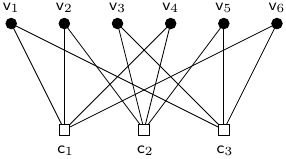}}
\vspace{-2.1ex}
\caption{Bipartite graph corresponding to the assignment matrix in \eqref{eq:Matrix}.}
\label{fig:bipartitegraph}
\vspace{-2.7ex}
\end{figure}

The work \cite{KEK2019R} introduced a quantitative group testing scheme based on regular GLDPC codes where the test matrix $\bA$ corresponds to the partity-check matrix of a GLDPC code. Particularly, the construction in \cite{KEK2019R} is as follows. Consider a regular $(\dv,\dc)$ bipartite graph with $n$ VNs and $m_\msfB$ CNs and its corresponding $m_\msfB\times n$ adjacency matrix $\bB$. 
To construct the test matrix $\bA$, each of the $\dc$ non-zero elements in a row of $\bB$ is replaced by a column of an $\ns\times \dc$ signature matrix $\bU=\left( \oneb^\top, \bH\transpose\right)\transpose$, where $\oneb$ is a $1\times \dc$ all-ones vector and $\bH$, of dimensions $t\log_2(\dc+1)\times n$, is the parity-check matrix of a $t$-error correcting BCH code of length $\dc$. Hence, $\ns=t\log_2(\dc+1)+1$, and the total number of tests is given by $m=m_\msfB \ns$.  (Note that, for a GLDPC code-based GT scheme, contrary to the bipartite graph in Fig.~\ref{fig:bipartitegraph}, each of the CNs corresponds to a bundle of $\ns$ tests.)

We denote by $\rate$ the \emph{rate} of the GT scheme, i.e., the ratio between the number of tests and the number of items.\footnote{Note that, interpreting $\bA$ as the parity-check matrix of a code, $\Omega=1-R$, where $R$ is the code rate.} For the construction in \cite{KEK2019}, 
\begin{align}\label{rate}
  \rate&=\frac{m}{n}=\frac{\dv}{\dc}\Big(t\ceil{\log_2(\dc+1)}+1\Big)\,,
\end{align}
where the ceiling function $\ceil{.}$ takes care of cases where $\dc+1$ is not a power of two.

Decoding to recover $\bx$ is performed via peeling decoding, where at each iteration, due to the $t$-error correcting capability of the BCH codes, a CN connected to $t$ or less unresolved defective items can identify them and their adjacent edges are peeled off the graph.

The probability that a defective item remains
unidentified over  iterations can be tracked via density evolution. Let $p\pwl$ the probability that a defective item remains unidentified at iteration $\ell$ and $q\pwl$  the probability that a  CN is resolved at iteration  $\ell$. The quantities $p\pwl$ and $q\pwl$ are given by the following density evolution equations \cite{KEK2019}, 
\begin{align*}
    q\pwl=&\sum_{i=0}^{t-1}\binom{\dc-1}{i}\left(p\pwlp\right)^i\left(1-p\pwlp\right)^{\dc-1-i}\\
     p\pwl=&\gamma \, (1-q\pwlp)^{\dv-1}\,.
\end{align*}

\section{Quantitative Group testing\\Based on LDPC codes}
In this section, we propose a novel GT scheme based on LDPC codes in which the test matrix $\bA$ is the parity-check matrix of  an LDPC code or, correspondingly is obtained from the bipartite graph of an LDPC code (as the one in Fig.~\ref{fig:bipartitegraph}). 
\subsection{Proposed GT scheme}
We consider a regular $(\dv,\dc)$ bipartite graph, where each VN  is connected to  $\dv$ CNs and each CN is connected to $\dc$ VNs.
The rate of the LDPC code-based GT scheme is
    $\rate=\frac{\dv}{\dc}\,,$
which can also be obtained from \eqref{rate} by setting $t=0$.

Similar to LDPC codes over the binary erasure channel and GLDPC code-based GT, decoding can be performed via peeling decoding. Peeling decoding gives rise to a sequence of residual graphs. Decoding is successful if eventually the decoder manages to peel off all VNs from the original graph, resulting in an empty graph.

Let $\dcell$ the degree of a generic CN $\msfc$ at iteration $\ell$ and $\bsell$ the corresponding syndrome (after the contribution of the resolved items in previous iterations has been removed). The proposed peeling decoding algorithm is based on the following observation: If $\bsell=0$, then all VNs connected to $\msfc$ are non-defective and can be resolved. Furthermore,  if  $\bsell=\dcell$, then all VNs connected to $\msfc$ are defective and can also be resolved. Otherwise, none of the connected VNs can be resolved by considering $\msfc$. This observation yields to the following peeling decoding algorithm:

\begin{enumerate}
\item Initialization: Set $s_i^{(\ell)}=s_i$ and all items as \emph{unresolved}
\item For $\ell\ge1$: For each CN $\msfc_i$ in the residual graph at iteration $\ell$,
\begin{itemize}
\item If $s_i^{(\ell)}=0$ declare all connected VNs as non-defective items and peel off their adjacent edges
\item If $s_i^{(\ell)}=d_{\msfc_i}^{(\ell)}$ declare all connected VNs as defective items, subtract $1$ from the syndrome of their neighboring CNs, and peel off their adjacent edges
\end{itemize}
\item If the resulting residual graph is empty or no edges have been peeled off in Step 2 (i.e., decoding stalls), stop the decoding. Otherwise, increase $\ell$ and return to 2)
\end{enumerate}

\vspace*{-2mm}
\subsection{Density Evolution}
\label{sec:DELDPC}

In this section, we derive the density evolution equations of the peeling decoding algorithm introduced in this section. For convenience, we group the VNs into two classes, the class of VNs corresponding to defective items, which we call \emph{defective} VNs, and the class of VNs corresponding to non-defective items, which we call \emph{non-defective} VNs.

  Let $p_0\pwl$ be the probability that a message from a non-defective VN  to a CN at iteration $\ell$ is an \emph{unresolved} message, and  $p_1\pwl$ the probability that a message from a defective VN  to a CN at iteration $\ell$ is \emph{unresolved}.
Also let $\qz$ be the probability that a message from a CN to a non-defective VN is a \emph{resolved} message, and $\qd$ be the probability that a message from a CN to a defective VN
 is \emph{resolved}. 
 
 \begin{proposition}
 The quantities $\pz$, $ \pd$, $\qz$, and $\qd$ are given by the following density evolution equations:
\begin{align}
\label{DEqz}
 \qz=&\sum_{i=0}^{\dc-1}\binom{\dc-1}{i} \ga^i(1-\ga)^{\dc-1-i} \left(1-\pdm\right)^i\\
 \label{DEqd}
  \qd=&\sum_{i=0}^{\dc-1}\binom{\dc-1}{i} \ga^i(1-\ga)^{\dc-1-i} \left(1-\pzm\right)^{\dc-1-i}\\
  \label{DEpz}
  \pz=&\left(1-\qzm\right)^{\dv-1}\\
  \pd=&\left(1-\qdm\right)^{\dv-1}\,.
  \label{DEpd}
\end{align}
 \end{proposition}
\begin{IEEEproof}
 The probability that $i$ out of the $\dc-1$ VNs connected to CN
 through its adjacent edges except the one on which the outgoing message is sent are defective, is given by a binomial distribution with parameters $\dc-1$ and $\gamma$, $\Bino(\dc-1,\ga)$.

A message from a CN $\msfc$ to a non-defective VN is resolved if all incoming messages from defective VNs are resolved or all VNs connected to $\msfc$ are non-defective. If the number of defective items connected to $\msfc$ is $i$, then, this occurs with probability $(1-p_1^{(\ell-1)})^{i}$. Considering that $i$ is binomially distributed and summing over all $i$, we obtain \eqref{DEqz}.
 
 Similarly,  a message from a CN $\msfc$ to a defective VN is resolved if all incoming messages from non-defective VNs are resolved or all VNs connected to $\msfc$ are defective~(i.e,  $i=\dc-1$), yielding  \eqref{DEqd}.
Finally, a message from a non-defective or defective VN to a CN is unresolved if all its incoming $\dc-1$ messages are unresolved, yielding    \eqref{DEpz} and   \eqref{DEpd}.
\end{IEEEproof}

\section{Group Testing with Spatial coupling}

In this section, we apply the concept of spatial coupling to the LDPC code-based GT scheme introduced in the previous section and the GLDPC code-based GT proposed in \cite{KEK2019R}.

\subsection{Group Testing based on Spatially-Coupled LDPC Codes}
\label{sec:GTSCLDPC}

Similar to SC-LDPC codes, the Tanner graph of a \emph{terminated} SC-LDPC code-based GT is constructed by placing $L$ copies of the bipartite graph of a $(\dv,\dc)$-regular LDPC code-based GT in $L$ spatial positions, each consisting of $n_\mathsf{b}$ VNs and $m$ CNs. We refer to $L$ as the coupling length and to $n_\mathsf{b}$ as the  component code block length. The $L$ copies are then coupled as follows: each VN at spatial position $\tau\in[L]$ is connected to $\dv$ CNs at positions in the range $[\tau,\tau+w]$, where $w$ is referred to as the coupling memory. For each connection, the position of the CN is uniformly and independently chosen from that range. Further, each CN at spatial position $\tau\in[L]$ is connected to $\dc$ CNs at positions in the range $[\tau,\tau-w]$.

As for SC-LDPC codes, the lower degree of the CNs at the boundaries of the coupled chain yield to a wave-like
decoding effect where a decoding wave propagates from the boundaries of the chain inward.

The rate of the SC-LDPC code-based GT scheme is 
\begin{align} 
\label{eq:rateSC}
\rate_\mathsf{SC}=\left(1+\frac{w}{L}\right)\rate\,,
\end{align}
with $\Omega=\frac{\dv}{\dc}$. Note that coupling implies an increase in the number of tests by a factor of $\frac{w}{L}$ compared to the uncoupled case\textemdash akin to the rate loss of SC-LDPC codes\textemdash that vanishes as $L$ becomes large.

The density evolution equations for SC-LDPC code-based GT are given in the following proposition.

\begin{proposition}
 The quantities $\pzt$, $ \pdt$, $\qzt$, and $\qdt$ are given by the following density evolution equations:
\begin{align*}
    \qzt=&\frac{1}{\w+1}\sum_{j=0}^\w\sum_{i=0}^{\dc-1}\Bino(\dc-1,i,\ga) \left(1-\pdtm\right)^i\\
    \qdt=&\frac{1}{\w+1}\sum_{j=0}^\w\sum_{i=0}^{\dc-1}\Bino(\dc-1,i,\ga) \left(1-\pztm\right)^{\dc-1-i}\\
  \pzt=&\frac{1}{\w+1}\sum_{j=0}^\w\left(1-\qztm\right)^{\dv-1}\\
  \pdt=&\frac{1}{\w+1}\sum_{j=0}^\w\left(1-\qdtm\right)^{\dv-1}\,.
\end{align*}
\end{proposition}

\subsection{Group Testing based on Spatially-Coupled GLDPC Codes}
\label{sec:GTSCGLDPC}

The coupling of GLDPC code-based GT is performed in a similar way as for SC-LDPC code-based GT. However, contrary to 
SC-LDPC code-based GT, which is obtained by coupling the bipartite graph corresponding to the test matrix $\bA$, the coupling of 
GLDPC code-based GT
 is performed over the bipartite  graph corresponding to the adjacency matrix $\bB$.

The rate of the SC-GLDPC code-based GT scheme is also given by  \eqref{eq:rateSC}, with $\Omega$ given in \eqref{rate}.

The density evolution equations for SC-GLDPC code-based GT are given in the following proposition.

\begin{proposition}
 The quantities $q\pwl_\tau$, and $p\pwl_\tau$ are given by the following density evolution equations:
\begin{align*}
    q\pwl_\tau=&\frac{1}{\w+1}\sum_{j=0}^\w\sum_{i=0}^{t-1}\binom{\dc-1}{i}\left(p\pwlp_{\tau-j}\right)^i\left(1-p\pwlp_{\tau-j}\right)^{\dc-1-i}\\
    p\pwl_\tau=&\frac{1}{\w+1}\sum_{j=0}^{\w}\gamma \left(1-q\pwlp_{\tau+j}\right)^{\dv-1}\,.
\end{align*}
\end{proposition}

\section{Numerical Results}

\subsection{Density Evolution Thresholds}

The density evolution equations derived in Sections~\ref{sec:DELDPC}, \ref{sec:GTSCLDPC}, and \ref{sec:GTSCGLDPC} can be used to analyze the behavior of GT in the limit of large $n$, and more precisely  to compute the GT \emph{threshold}. 
In particular, for a fixed prevalence  $\ga$, the GT threshold $\Omegath$ is defined as the minimum rate\textemdash the minimum number of tests per item\textemdash required for perfect detection of the defective items. Conversely,  for a fixed rate $\Omega$, the GT threshold $\gammath$ is defined as the maximum prevalence that allows perfect detection of the defective items.\footnote{We consider the threshold $\Omegath$, as this is the quantity considered in \cite{KEK2019R}. However, from a coding perspective, it is interesting to fix the rate of the scheme and compute $\gammath$, which is  akin to the belief propagation threshold for codes-on-graphs.}

Here, we give density evolution results for the proposed LDPC code-based and spatially-coupled GT schemes 
and compare them with the GLDPC code-based scheme in \cite{KEK2019R}. 

\begin{table}[!t]
\caption{$\Omegath$ for  $\ga=0.15\%$ with GLDPC code-based group testing}
\vspace{-3.5ex}
\begin{center}
\begin{tabular}{ccccccc}
\toprule
{}&{}&\multicolumn{5}{c}{coupling memory}\\

$t$&$\dv$&$\w=0$&$\w=1$&$\w=2$&$\w=5$&$\w=10$\\[0.5mm]
\otoprule
\multirow{3}*{$1$}    &$2$&3.3588 &3.3574&3.3564  &3.3564&3.3564\\[0.5mm]
                      &$3$&2.2374 &1.9968&1.9956& 1.9956&1.9956\\[0.5mm]
                      &$4$&2.3715 &2.0432& 2.0328& 2.0320 & 2.0320\\[0.5mm]
\hline
\multirow{3}*{$2$}    &$2$&2.2472 &2.1286 &2.1277 &2.1268&2.1268\\[0.5mm]
                      &$3$&2.4574 &1.9506&1.9310  &1.9310&1.9310\\[0.5mm]
                      &$4$&2.8612 &2.1726 &2.1268  &2.0650&2.0650\\[0.5mm]
\hline
\multirow{3}*{$3$} & 2 & 2.1926 & 1.9655 & 1.9639 & 1.9629 & 1.9623 \\[0.5mm]
        & 3 & 2.7106 & 2.1056 & 2.0443 & 2.0415 & 2.0408 \\[0.5mm]
        & 4 & 3.3713 & 2.2504 & 2.0637 & 2.0369 & 2.0364 \\[0.5mm]
\hline
\multirow{3}*{$5$}    &$2$&2.4079 &2.0580&2.0367  &2.0364&2.0364\\[0.5mm]
                      &$3$&3.0622 &2.3407&2.1884  &2.1686&2.1686\\[0.5mm]
                      &$4$&3.7795 &2.2653 &2.2655 &2.1691&2.1691\\[0.5mm]
\bottomrule                      
\end{tabular} 
\end{center}
\label{TablechangingR_GLDPC}
\vspace{-3.7ex}
\end{table}

 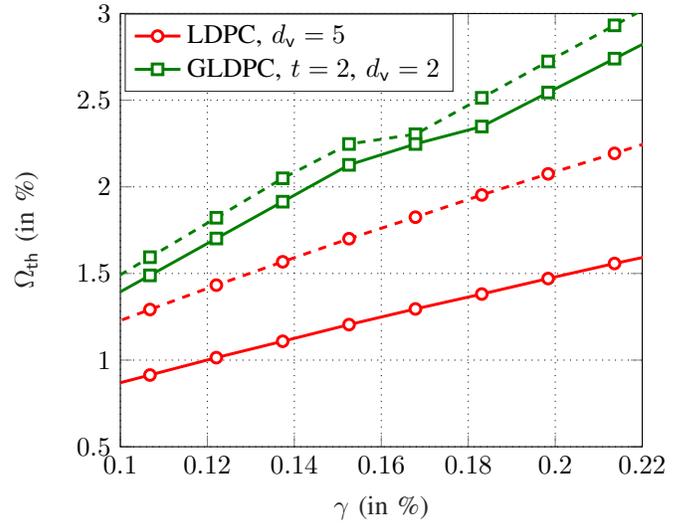
\begin{figure}[t]
\centering
 \resizebox{!}{0.8\linewidth}{
%
%
\begin{tikzpicture}

\begin{axis}[%
xmin=0.1,
xmax=0.22,
xlabel style={font=\color{white!15!black}},
xlabel={$\ga$ (in \%) },
ymin=0.5,
ymax=3,
 xtick={0.10, 0.12, 0.14, 0.16, 0.18, 0.20, 0.22},
ylabel style={font=\color{white!15!black}},
ylabel={$\Omegath$ (in \%)},
axis background/.style={fill=white},
xmajorgrids,
ymajorgrids,
grid style={dotted,draw=black!90},
legend style={at={(axis cs: 0.101,3.00)}, anchor=north west, legend cell align=left, align=left, draw=white!15!black}
]

\addplot [color=red, dashed, line width=1.1pt, mark=*, mark options={solid, fill=white}, forget plot]
  table[row sep=crcr]{%
0.0762939453125	0.998003992015968\\
0.091552734375	1.14942528735632\\
0.1068115234375	1.29198966408269\\
0.1220703125	1.43266475644699\\
0.1373291015625	1.56739811912225\\
0.152587890625	1.70068027210884\\
0.1678466796875	1.82481751824817\\
0.18310546875	1.953125\\
0.1983642578125	2.0746887966805\\
0.213623046875	2.19298245614035\\
0.2288818359375	2.31481481481482\\
};

\addplot [color=red, solid,  line width=1.1pt, mark=*, mark options={solid, fill=white}]
  table[row sep=crcr]{%
0.0762939453125	0.7061767578125\\
0.091552734375	0.814361572265625\\
0.1068115234375	0.91400146484375\\
0.1220703125	1.01425170898438\\
0.1373291015625	1.10870361328125\\
0.152587890625	1.204833984375\\
0.1678466796875	1.29531860351563\\
0.18310546875	1.3812255859375\\
0.1983642578125	1.47064208984375\\
0.213623046875	1.5576171875\\
0.2288818359375	1.63934426879883\\
};
\addlegendentry{LDPC, $\dv=5$}

\addplot [color=darkgreen, dashed, line width=1.1pt, mark=square*, mark options={solid, fill=white}, forget plot]
  table[row sep=crcr]{%
   0.076293945312500   1.229508196721312\\
   0.091552734375000   1.366120218579234\\
   0.106811523437500   1.593879502709595\\
   0.122070312500000   1.821493624772308\\
   0.137329101562500   2.049180327868851\\
   0.152587890625000   2.247191011235962\\
   0.167846679687500   2.303455182774155\\
   0.183105468750000   2.513661202185791\\
   0.198364257812500   2.723505032563645\\
   0.213623046875000   2.931803696622055\\
   0.228881835937500   3.142076502732239\\
};

\addplot [color=darkgreen, solid ,line width=1.1pt, mark=square*, mark options={solid, fill=white}]
  table[row sep=crcr]{%
0.0762939453125	1.148691767709\\
0.091552734375	1.27616130678918\\
0.1068115234375	1.48898153662895\\
0.1220703125	1.70183798502382\\
0.1373291015625	1.91424196018376\\
0.152587890625	2.12675457252232\\
0.1678466796875	2.24719101123596\\
0.18310546875	2.34813680449208\\
0.1983642578125	2.54424778761063\\
0.213623046875	2.73972602739726\\
0.2288818359375	2.93554562858966\\
};
\addlegendentry{GLDPC, $t=2$, $\dv=2$}

\end{axis}

\end{tikzpicture}%
}
\vspace{-5ex}
\caption{$\Omegath$ as a function of $\gamma$ for LDPC code-based and GLDPC code-based schemes. Dashed lines are for the uncoupled schemes, while solid lines are for the coupled schemes.}
\label{Limit}
\vspace{-2ex}
\end{figure}

In Table~\ref{TablechangingR_GLDPC},  we give $\Omegath$ for a prevalence $\ga=100/2^{16}$ for GLDPC code-based GT.\footnote{We use this prevalence as it is the one considered in \cite[Fig.~2]{KEK2019R}} The uncoupled case, $w=0$, corresponds to the scheme in  \cite{KEK2019R}. We can see that coupling improves the threshold $\rate$ (except for $t=1$ and $\dv=2$), and the improvement increases with increasing $t$ and $\dv$. For both the uncoupled and coupled cases, the best threshold is obtained for $t=3$ and $\dv=2$.

In Fig.~\ref{Limit}, we plot threshold $\Omegath$ as a function of the prevalence $\gamma$ (both in percentage) for the proposed LDPC code-based GT scheme with $\dv=5$,
the  GLDPC code-based GT scheme of \cite{KEK2019R} with $t=2$ and $\dv=2$,   and the coupled versions of both. We observe that the LDPC code-based GT scheme significantly outperforms the scheme in  \cite{KEK2019R}. Spatial coupling  improves $\Omegath$ for both schemes, with the largest improvement for the proposed LDPC code-based GT scheme.

Finally, in Tables~\ref{TableFIXR_GLDPC} and~\ref{TableFIXR_LDPC}, we give $\gammath$ for $\rate=5\%$ for GLDPC code-based and LDPC code-based GT, respectively. For  LDPC code-based GT (Table~\ref{TableFIXR_LDPC}), we observe that with coupling the threshold improves with increasing $\dv$ (similarly to LDPC codes). Compared to GLDPC code-based GT (Table~\ref{TableFIXR_GLDPC}), LDPC code-based GT achieves significantly higher thresholds.  Furthermore, we generally observe that the thresholds tend to converge to a constant value for a large enough coupling memory $w$. 


\begin{table}[!t]
\caption{$\gamma_\mathrm{th}$ for $\rate=5\%$ with GLDPC Code-Based Group Testing}
\vspace{-3.5ex}
\begin{center}
\begin{tabular}{ccccccc}
\toprule
{}&{}&\multicolumn{5}{c}{coupling memory}\\

$t$&$\dv$&$\w=0$&$\w=1$&$\w=2$&$\w=5$&$\w=10$\\[0.5mm]
\otoprule
\multirow{3}*{$1$} &2&0.2487 & 0.2502 & 0.2502 & 0.2502 &0.2502\\[0.5mm]
                    &3  &0.3708 & 0.4166 & 0.4166 & 0.4166 & 0.4166\\[0.5mm]
                    &4 &0.3510 & 0.4395 & 0.4425 & 0.4425 & 0.4425\\[0.5mm]
\hline
\multirow{3}*{$2$}      & 2 & 0.3983 & 0.4257 & 0.4257 & 0.4257 & 0.4257\\[0.5mm]
                        & 3 & 0.3372 & 0.4242 & 0.4288 & 0.4288 & 0.4288\\[0.5mm]
                        & 4 & 0.2884 & 0.4120 & 0.4318 & 0.4333 & 0.4333\\[0.5mm]
\hline
\multirow{3}*{$3$}  &    2 & 0.3784 & 0.4211 & 0.4227 & 0.4227 & 0.4227 \\[0.5mm]
        &3 & 0.3189 & 0.4257 & 0.4379 & 0.4379 & 0.4395 \\[0.5mm]
        &4 & 0.2441 & 0.3662 & 0.3983 & 0.4028 & 0.4028 \\[0.5mm]
\hline
\multirow{3}*{$5$}&    2 & 0.3418 & 0.3998 & 0.4044 & 0.4044 & 0.4044 \\[0.5mm]
       & 3 & 0.2686 & 0.3784 & 0.4044 & 0.4089 & 0.4089 \\[0.5mm]
        &4 & 0.2014 & 0.3159 & 0.3616 & 0.3769 & 0.3769 \\[0.5mm]
\bottomrule                      
\end{tabular} 
\end{center}
\label{TableFIXR_GLDPC}
\vspace{-4.5ex}
\end{table}
\begin{table}[!t]
\caption{$\gamma_\mathrm{th}$ for $\rate=5\%$ with LDPC Code-Based Group Testing}
\vspace{-3.5ex}
\begin{center}
\begin{tabular}{cccccc}
\toprule
{}&\multicolumn{5}{c}{coupling memory}\\

$\dv$&$\w=0$&$\w=1$&$\w=2$&$\w=5$&$\w=10$\\[0.5mm]
\otoprule
                    3   &0.4555 & 0.5544 & 0.5508 & 0.5559 &0.5559\\[0.5mm]
                    4   &0.5982 & 0.8423 & 0.8532 & 0.8540 & 0.8540\\[0.5mm]
                    5   &0.6416 & 0.9682 & 1.0270 & 1.0274 &1.025\\[0.5mm]
                    6   &0.6464 & 1.0044 & 1.1196 & 1.1325 &1.1327\\[0.5mm]
                    7   &0.6353 & 0.9999 & 1.1585 & 1.1978 &1.1980\\[0.5mm]
                    10  &0.5773 & 0.9188 & 1.1272 & 1.2814 &1.2816\\[0.5mm]
\bottomrule                      
\end{tabular} 
\end{center}
\label{TableFIXR_LDPC}
\vspace{-3.8ex}
\end{table}

\subsection{Simulation Results}\label{sec:simulation}

\begin{figure}[t]
\centering
 \resizebox{!}{0.75\linewidth}{
%
%
\definecolor{mycolor1}{rgb}{0.00000,0.44700,0.74100}%
\definecolor{mycolor2}{rgb}{0.85000,0.32500,0.09800}%
\definecolor{mycolor3}{rgb}{0.92900,0.69400,0.12500}%
\definecolor{mycolor4}{rgb}{0.49400,0.18400,0.55600}%
\begin{tikzpicture}

\begin{axis}[%
width=3.2in,
height=2.79in,
at={(0.61in,0.633in)},
scale only axis,
xmin=0.2,
xmax=1.4,
xlabel style={font=\color{white!15!black}},
xlabel={$\ga$ [\%]},
ymode=log,
ymin=1e-05,
ymax=1,
yminorticks=true,
ylabel style={font=\color{white!15!black}},
ylabel={misdetection rate},
axis background/.style={fill=white},
grid style={dotted,draw=black!90},
xmajorgrids,
ymajorgrids,
yminorgrids,
legend style={at={(axis cs: 0.68,0.000012)}, anchor=south west, legend cell align=left, align=left, draw=white!15!black}
]
\addplot [color=darkblue,  solid, line width=1.1pt,  mark=square*, mark options = {fill = white}, mark size=2pt]
  table[row sep=crcr]{%
0.6	0.959629224072652\\
0.558	0.921751266547778\\
0.55	0.843954934762279\\
0.545	0.742470141509151\\
0.54	0.506514931101383\\
0.53	0.0155947813194082\\
0.5285	8.425123462e-05\\
};
\addlegendentry{$\dv=3$}

\addplot [color=darkblue,  dashed, line width=1.1pt,forget plot, mark=square*, mark options = {solid,fill = white}, mark size=2pt]
  table[row sep=crcr]{%
0.8	0.999616498714863\\
0.7	0.997300919461865\\
0.642	0.993030121643146\\
0.613	0.98847549035344\\
0.59	0.982104151614517\\
0.578	0.978884152329271\\
0.571	0.976022558895487\\
0.568	0.976040447598568\\
0.56	0.973335194750803\\
0.55	0.968338476760348\\
0.54	0.958842616311693\\
0.48	0.835059382422803\\
0.43	0.113533875767553\\
0.4212	0.0523293263074571\\
0.4132	0.0261993600881066\\
0.4052	0.00670970029575547\\
0.3902	0.00074334690145118\\
};

\addplot [darkgreen, solid, line width=1.1pt, mark=triangle*, mark options={fill=white} ,mark size=3pt]
 table[row sep=crcr]{%
1.2	0.994031892000529\\
1.1	0.988397278746431\\
1.05 0.972703173704788\\
1.02 0.863246106092585\\
1	0.0929694677827095\\
0.992	0.00539447124347729\\
0.99	0.000315120564215646\\
};
\addlegendentry{$\dv=5$}

\addplot [darkgreen, dashed, line width=1.1pt, forget plot, mark=triangle*, mark options={solid,fill=white} ,mark size=3pt]
  table[row sep=crcr]{%
0.8	0.999141518612694\\
0.7	0.987667805597565\\
0.65	0.69162681348712\\
0.642	0.505713319464229\\
0.63	0.34061740378473\\
0.62	0.220046856202115\\
0.613	0.133956705083763\\
0.611	0.11082768987587\\
0.6	0.0426049490982575\\
0.59	0.0132374202897586\\
0.58	0.00296342702870806\\
0.5795	0.00309401819386547\\
0.579	0\\
};

\addplot [red, thick, line width=1.1pt, mark=diamond*, mark options={solid,fill=white}, mark size=3pt]
  table[row sep=crcr]{%
1.3	0.94162893041651\\
1.28	0.92162893041651\\
1.26	0.81510924849125\\
1.23	0.0195823378292815\\
1.22	0.00015512234445\\
};
\addlegendentry{$\dv=10$}

\addplot [red, thick, line width=1.1pt, dashed, mark=diamond*, mark options={solid,fill=white}, mark size=3pt]
  table[row sep=crcr]{%
0.7	0.971234870708294\\
0.6	0.937364870708294\\
0.58	0.587300337457818\\
0.57	0.436926355834126\\
0.565	0.378719868917531\\
0.56	0.292840876743524\\
0.55	0.135349554183651\\
0.535	0.0368211620189647\\
0.532	0.0264860192524994\\
0.529	0.0116895710447485\\
0.51  8.415628329330000e-05\\
};

\addplot [color=darkblue,  solid, line width=1.1pt, forget plot,  mark=square*, mark options = {fill = white}, mark size=2pt]
  table[row sep=crcr]{%
0.6	0.959629224072652\\
0.56	0.916752543461443\\
0.558	0.921751266547778\\
0.557	0.909091574169722\\
0.555	0.908739897881317\\
0.55	0.843954934762279\\
0.545	0.742470141509151\\
0.54	0.506514931101383\\
0.53	0.0155947813194082\\
0.5285	8.425123462e-05\\
};
\addplot [darkblue, dashed, line width=1.1pt, forget plot]
  table[row sep=crcr]{%
0.4513	10e-6\\
0.4513	2e-4\\
};

\addplot [darkblue, solid, line width=1.1pt, forget plot]
  table[row sep=crcr]{%
0.5559	10e-6\\
0.5559	2e-4\\
};

\addplot [darkgreen, dashed, line width=1.1pt, forget plot]
  table[row sep=crcr]{%
0.6394	10e-6\\
0.6394	2e-4\\
};

\addplot [darkgreen, solid, line width=1.1pt, forget plot]
  table[row sep=crcr]{%
1.0274	10e-6\\
1.0274	2e-4\\
};

\addplot [red, dashed, line width=1.1pt, forget plot]
  table[row sep=crcr]{%
0.5773	10e-6\\
0.5773	2e-4\\
};

\addplot [red, solid, line width=1.1pt, forget plot]
  table[row sep=crcr]{%
1.2814	10e-6\\
1.2814	2e-4\\
};

\end{axis}

\end{tikzpicture}%
}
\vspace{-2ex}
\caption{Misdetection rate for uncoupled (dashed) and coupled (solid) LDPC code-based GT.}
\label{SimulationLDPC}
\vspace{-3ex}
\end{figure}
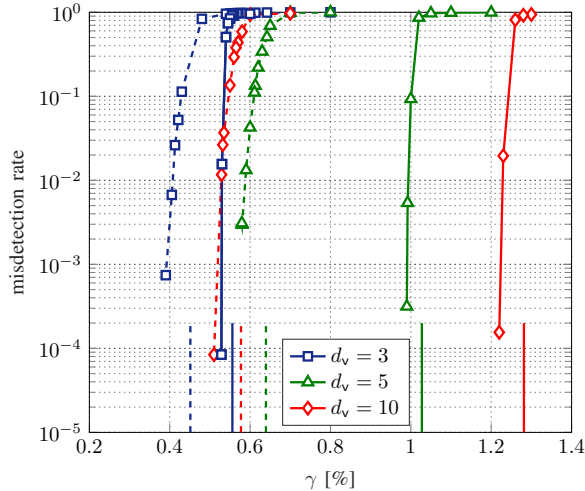

In this section, we give simulation results for finite block length.
In Fig.~\ref{SimulationLDPC}, we plot the misdetection rate, i.e., the fraction of defective items not identified, as a function of the prevalence $\gamma$ for the proposed LDPC code-based GT scheme with no coupling~(dashed lines) and coupling~(solid lines) for $\dv=3$, $5$, and $10$, and rate $\Omega=0.05$. The block length of the uncoupled scheme is $n=153000$. For the coupled scheme, we consider $\w=5,L=200$, and component code block length $n_{\mathsf{b}}=102000$. Further, the coupled scheme is decoded by iterating on the entire chain~(not using a window decoder).
We observe that coupling significantly improves performance, particularly for large $\dv$, in agreement with the density evolution results (cf. Table~\ref{TableFIXR_LDPC}). The density evolution thresholds are given by the vertical lines. We remark that the latency (defined as the number of items that need to be tested before delivering test results) of the coupled scheme is much larger than that of the uncoupled scheme. The former is $n_{\mathsf{b}}L$, while the latter is $n$. Note, however, that increasing $n$ for the uncoupled scheme marginally improves its performance (the limit is given by the density evolution threshold), hence the figure highlights how much one can gain with coupling if latency is not a problem.

In Fig.~\ref{SimulationGLDPC_FX}, we plot the misdetection rate for uncoupled (dashed lines) and coupled (solid lines) GLDPC code-based GT with $t=3$, $\dv=3$, and  $\rate=0.05$. Further, we consider two component code block lengths,   $n_\mathsf{b}=153000$ and $n_\mathsf{b}=10200$. For $n_\mathsf{b}=153000$, we assume full decoding, while for $n_\mathsf{b}=10200$
we assume a sliding window (SW) decoding \cite{Iye12} with window size $W=15$. The latency of the coupled scheme is therefore $n_\mathsf{b}W$, i.e.,  the latencies of the uncoupled scheme and coupled scheme with SW decoding are identical.  Notably, for SW decoding and  the same latency, coupling still outperforms the uncoupled scheme.  As for the LDPC code-based GT scheme, coupling with full decoding significantly improves performance.

In Fig.~\ref{Sim_GLDPC_VarRate}, we consider the scenario with fixed prevalence $\gamma$ and varying rate $\Omega$, corresponding to \cite[Fig.~2]{KEK2019R}. For GLDPC code-based GT with $t=2$ and $d_v=2$, we plot the misdetection rate as a function of the number of tests per defective item. Each point in this plot corresponds to a different rate $\Omega$. We observe for the uncoupled scheme (dashed) that the curve flattens as the  number of tests increases, while it decays very steeply for the coupled scheme (solid).

\section{Conclusion}\label{sec:Conclusion}
Our analysis demonstrates that spatial coupling  improves the performance of GLDPC code-based and LDPC code-based  GT schemes in terms of both the asymptotic performance and the error floor.
Our numerical results indicate that the thresholds tend to converge to a constant value as $w$ increases. We plan to investigate threshold saturation in an extended version of this paper. Further, both the thresholds and the finite length simulations show that our proposed LDPC code-based scheme performs significantly better than the GLDPC code-based schemes from the literature, based on $t$-error correcting BCH codes, both with and without spatial coupling.

 \begin{figure}[t]
\centering
 \resizebox{!}{0.68\linewidth}{
%
%
\begin{tikzpicture}

\begin{axis}[%
xmin=0,
xmax=0.6,
xlabel style={font=\color{white!15!black}},
xlabel={$\ga$ [\%]},
ymode=log,
ymin=1e-06,
ymax=1.04084213360753,
yminorticks=true,
ylabel style={font=\color{white!15!black}},
ylabel={misdetection rate},
axis background/.style={fill=white},
grid style={dotted,draw=black!90},
xmajorgrids,
ymajorgrids,
yminorgrids,
legend style={at={(axis cs: 0.001,1.1e-6)}, anchor=south west, legend cell align=left, align=left, draw=white!15!black}
]
\addplot [color=darkblue,  solid, line width=1.1pt, mark=square*, mark options = {solid,fill = white}, mark size=2pt]
  table[row sep=crcr]{%
0.6122	0.998679432155827\\
0.4132	0.965774735532047\\
0.3932	0.950892126370928\\
0.3902	0.94284251834158\\
0.3872	0.942440006712536\\
0.3812	0.950076674050094\\
0.3521	0.899420452421013\\
0.3195	0.476185564883764\\
0.3105	0.353044854881266\\
0.302	0.189078453901017\\
0.294	0.0625347229938443\\
0.2795	0.00993580206910196\\
0.2775	0.0105432690419138\\
0.2765	0.00692063841670141\\
0.275	0.00662387033694582\\
0.2705	0.00107213795594077\\
};
\addlegendentry{uncoupled}

\addplot [color=darkgreen,  dashed, line width=1.1pt,  mark=triangle*, mark options = {fill = white}, mark size=2pt]
  table[row sep=crcr]{%
0.6122	0.990116600110438\\
0.4132	0.918452500561459\\
0.3932	0.852349669007526\\
0.3902	0.852071452849428\\
0.3872	0.809681302667427\\
0.3812	0.832058311294588\\
0.3521	0.342661423668524\\
0.3195	0.0105928725188298\\
0.3105	0.00053332071035794\\
0.302	0\\
0.294	1.67007361684503e-06\\
0.275	0\\
};
\addlegendentry{coupled, SW}

\addplot [color=darkgreen,  solid, line width=1.1pt, mark=triangle*, mark options = {solid,fill = white}, mark size=2pt]
  table[row sep=crcr]{%
0.8	0.999642731331463\\
0.7	0.997590626594281\\
0.655	0.994130920643521\\
0.58	0.929220534116504\\
0.464	0.61582744266944\\
0.455	0.201895778232058\\
0.445	0.0224585981546525\\
0.435	0.000251507271918465\\
0.430	0\\
};
\addlegendentry{coupled}

\end{axis}

\end{tikzpicture}%
}
\vspace{-3ex}
\caption{Misdetection rate for uncoupled (dashed) and coupled (solid) GLDPC code-based GT with $t=3$ and $\dv=3$.}
\label{SimulationGLDPC_FX}
\vspace{-1.0ex}
\end{figure}
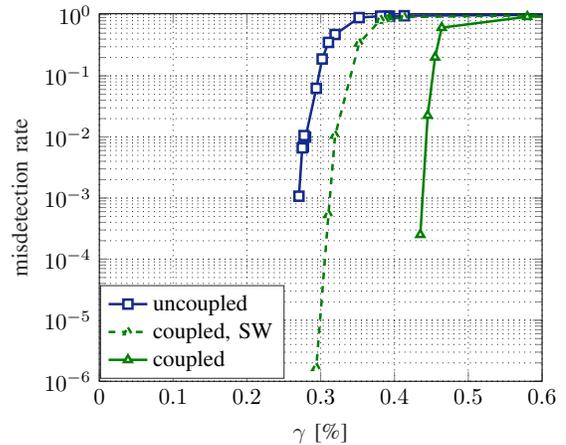
 \begin{figure}[t]
\centering
 \resizebox{!}{0.68\linewidth}{
%
%
\definecolor{mycolor1}{rgb}{0.00000,0.44700,0.74100}%
\definecolor{mycolor2}{rgb}{0.85000,0.32500,0.09800}%
\definecolor{mycolor3}{rgb}{0.92900,0.69400,0.12500}%
\definecolor{mycolor4}{rgb}{0.49400,0.18400,0.55600}%
\begin{tikzpicture}

\begin{axis}[%
xmin=0,
xmax=60,
xlabel style={font=\color{white!15!black}},
xlabel={number of tests per defective item},
ymode=log,
ymin=1e-06,
ymax=1,
yminorticks=true,
ylabel style={font=\color{white!15!black}},
ylabel={misdetection rate},
axis background/.style={fill=white},
grid style={dotted,draw=black!90},
xmajorgrids,
ymajorgrids,
yminorgrids,
legend style={at={(axis cs: 59,1.2e-6)}, anchor=south east, legend cell align=left, align=left, draw=white!15!black}
]
\addplot [color=darkblue,  solid, line width=1.1pt,  mark=square*, mark options = {solid,fill = white}, mark size=2pt]
  table[row sep=crcr]{%
6.49396201833943	0.996041489201784\\
10.00200040008	0.901330266053211\\
11.988011988012	0.756443556443556\\
13.9979283066106	0.524050440534801\\
16.9841973120662	0.043903451524187\\
17.986837725437	0.0168398509719309\\
20.6905651023133	0.00202607610929416\\
22.9861853026331	0.000990404766735192\\
26.7132038836428	0.00054813766237913\\
28.5270005259291	0.000361088611145175\\
31.1046971295208	0.000172937312225907\\
};
\addlegendentry{uncoupled}

\addplot [darkgreen, solid, line width=1.1pt, mark=triangle*, mark options={fill=white} ,mark size=2pt]
 table[row sep=crcr]{%
7.371	0.97937\\
11.55	0.81182\\
12.6	0.73458\\
13.65	0.62458\\
14.7	0.45865\\
15.4875	0.056203\\
15.6975	6.7741935483871e-05\\
};
\addlegendentry{coupled}

\end{axis}

\end{tikzpicture}%
}
\vspace{-3ex}
\caption{Misdetection rate as a function of the number of tests per defective item for GLDPC code-based GT with $t=2$ and $\dv=2$, and $\ga=0.15\%$. 
}
\label{Sim_GLDPC_VarRate}
\vspace{-2ex}
\end{figure}
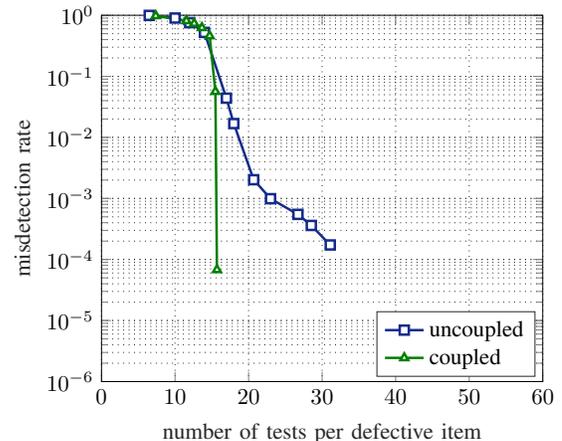


\newpage

\end{document}